\documentclass[final,5p,times,twocolumn,authoryear]{elsarticle}

\usepackage{amssymb}
\usepackage{amsmath}

\usepackage{lineno}

\usepackage{hyperref}

\journal{Applied Radiation and Isotopes}
\begin{document}

\begin{frontmatter}



\title{Development of a low background liquid scintillation counter for a shallow underground laboratory}


\author[PNNL,TAMU]{J.L.~Erchinger\corref{cor}}\ead{jennifer.erchinger@pnnl.gov}%\ead[url]{http://www.pnnl.gov/}
\author[PNNL]{C.E.~Aalseth}
%\author[PNNL]{T.A.~Beacham} % Pb-210 assay has been removed
\author[PNNL]{B.E.~Bernacki}
\author[PNNL]{M.~Douglas}
\author[PNNL]{E.S.~Fuller}
\author[PNNL]{M.E.~Keillor}
\author[PNNL]{S.M.~Morley}
\author[PNNL]{C.A.~Mullen}
\author[PNNL]{J.L.~Orrell}
\author[PNNL]{M.E.~Panisko}
%\author[PNNL]{S.M.~Shaff} % Pb-210 assay has been removed
\author[PNNL]{G.A.~Warren}
\author[PNNL,WU]{R.O.~Williams}
\author[PNNL]{M.E.~Wright}

\cortext[cor]{Corresponding author.}
\address[PNNL]{Pacific Northwest Laboratory, Richland, WA 99352, USA}
\address[TAMU]{Texas A\&M University, College Station, TX 77840, USA}
\address[WU]{Wittenberg University, Springfield, OH 45504, USA}

\begin{abstract}
Pacific Northwest National Laboratory has recently opened a shallow underground laboratory intended for measurement of low-concentration levels of radioactive isotopes in samples collected from the environment. The development of a low-background liquid scintillation counter is currently underway to further augment the measurement capabilities within this underground laboratory. Liquid scintillation counting is especially useful for measuring charged particle (e.g., $\beta$, $\alpha$) emitting isotopes with no (or very weak) gamma-ray yields. The combination of high-efficiency detection of charged particle emission in a liquid scintillation cocktail coupled with the low-background environment of an appropriately-designed shield located in a clean underground laboratory provides the opportunity for increased-sensitivity measurements of a range of isotopes. To take advantage of the 35 meters-water-equivalent overburden of the underground laboratory, a series of simulations have evaluated the scintillation counter's shield design requirements to assess the possible background rate achievable. This report presents the design and background evaluation for a shallow underground, low background liquid scintillation counter design for sample measurements.
\end{abstract}

\begin{keyword}
%% keywords here (maximum of 6), in the form: keyword \sep keyword



liquid scintillation counting \sep low background \sep shallow underground laboratory

\end{keyword}

%% Highlights
%%
%% Adding information on Highlights
%%
%% Highlights are a short collection of bullet points that convey the core findings of the article. Highlights are optional and should be submitted in a separate file in the online submission system. Please use 'Highlights' in the file name and include 3 to 5 bullet points (maximum 85 characters, including spaces, per bullet point). See http://www.elsevier.com/highlights for examples.
%%
%% #1 - Graded-shielding can produce an ultra-low-background liquid scintillation counter.
%% #2 - Location in a shallow underground cleanroom further enhances background reduction.
%% #3 - A novel light collection design and selected low background materials are utilized.
%% #4 - The background is predicted to be 10-100 times below typical commercial systems.
%% #5 - Simulations tentatively predict a background rate of order 10 counts per day.
%%

\end{frontmatter}

%\linenumbers

\section{Introduction}
\label{Introduction}

To further enhance the low-level radiation detection capabilities of the Pacific Northwest National Laboratory shallow underground laboratory (\cite{shallowlab}), the development of a low-background liquid scintillation counter is underway. The instrument is being developed to measure a range of low-level beta-emitting isotopes collected in the environment such as tritium, $^{14}$C, strontium isotopes, and others. Such measurements impact a range of sciences including studies of environmental transport mechanisms, biological radio-isotopic dating, radioactive fallout, or tracking of nuclear accident effluent (\cite{chapter9}).

\begin{figure*}[ht!]
\begin{center}
\includegraphics[width=2\columnwidth]{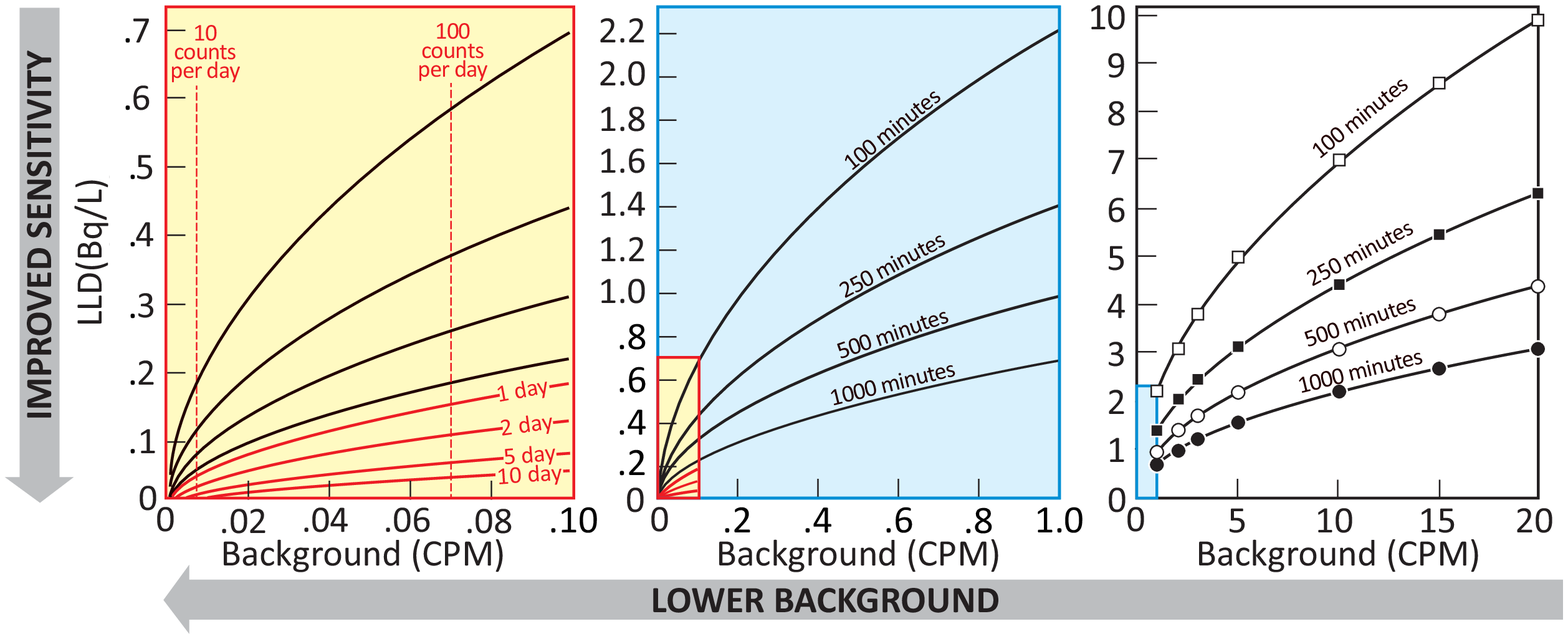}
\caption{\label{fig:lowbackground} A ``text book'' example of the relation between \textit{sensitivity} versus \textit{background \& count duration} with extrapolations to a shallow underground, low background liquid scintillation counter design. The ``text book'' example (right most panel) presents typical sensitivity levels for measurements of a 10~mL tritium sample given typical background rates and count duration and with 35\% counting efficiency (\cite{chapter7}). The panels to the left progressively reduce the background rate in the proposed instrument and show the relevant sensitivity levels attained for differing count duration. The count duration curves derived from the Currie formulation (Eq.~\ref{eq:toyCurrie}) are provided to guide the eye and may be impractical for some choices of background and count duration. In these figures the background rate is quantified in counts per minute (CPM) and the minimum detectable activity sensitivity is presented as a lower limit of detection (LLD) value.}
\end{center}
\end{figure*}

Liquid scintillation counting (LSC) identifies radioactive decay through detection of the ensuing scintillation photons produced by charged particles emitted into the scintillation cocktail. The photons are detected and this signal is amplified with one or more photomultiplier tubes (PMTs) for processing in a data acquisition system. Liquid scintillation counting provides a mechanism for counting $\beta$ and $\alpha$ emitters down to very low concentrations, due primarily to the high detection efficiency achievable through the use of a scintillation cocktail detection medium. Measurement of tritium, $^{3}$H, is useful for bench-marking a system's performance. Typical commercial low-background liquid scintillation counters can achieve detection limits on the order of 1~Bq of $^{3}$H per liter in the presence of $\sim$1~count per minute (cpm) backgrounds for 1000 minute (min) counting times. For the present study, consider the extrapolation of the decrease in lower limit of detection as a function of decreased background count rate, as shown in Figure~\ref{fig:lowbackground} for a series of counting times. The $^{3}$H counting efficiency in the extrapolated example is 35\% for a constant sample size of 10~mL (\cite{chapter7}). A combination of decreasing the background to the order of ten to one hundred counts per day and increasing the counting times to days results in a lower limit of detection more than an order of magnitude lower than typically achieved in commercial instruments.
%Added efficiency and sample size from Salonen Ch.7. JLE 02/27/15

%Changed figure reference to Salonen Ch7 from Ch9. JLE 02/27/15

The design presented in this report aims to reduce the background to a range from 0.01 to 0.07~cpm. In the case of tritium (See Fig.~\ref{fig:lowbackground}) this results in detection limits between 0.05 and 0.2~Bq/L depending on the number of days a sample is counted. These estimates employ Currie's equation for minimum detectable activity (MDA), given in Equation \ref{eq:toyCurrie} (\cite{currie}), where C$_{b}$ is the background count rate (cpm), T$_{b}$ is the background count time (min), $\varepsilon$ is the counting efficiency, V$_{s}$ is the sample weight (grams), and T$_{s}$ is the sample count time (min). 
\begin{equation}
\label{eq:toyCurrie}
\textrm{MDA} \equiv L_{D} = \frac{2.71 + 4.65\sqrt{C_{b}T_{b}}}{\varepsilon \times V_{s} \times T_{s} \times 60}
\end{equation}
The factor of 60 is used to express the MDA in terms of Bq/g. When the measurement is quantified for liquid volumes, the sample weight is replaced by a volume (milliliters, mL) resulting in a MDA reported in units of Bq/mL. The constant 2.71 is the frequently used value to account for a zero blank case corresponding to a 5\% probability of false negatives and 4.65 accounts for a 5\% probability of making Type~I or Type~II errors.  The Currie equation provides a consistent way of quantitatively evaluating measurements and is valuable for revealing the potential value of a custom low background LSC system. Foreshadowing the background estimates described in Section~\ref{sec:BackgroundEstimates}, the ultra-low-background liquid scintillation counter (ULB LSC) under development is expected to lower detection limits by more than a factor of 10 in comparison to commercial systems located in surface laboratories. In some cases, the result of the background reduction is used to directly improve the MDA of the measurement. In other cases, the 10-100 reduction factor in background rate is used as a trade-off with other factors in the Currie equation (such as sample mass or volume) to reach the same target MDA for a given measurement. For example, an improved MDA may allow greater reach of age-dating in a geochronology scenario while a fixed MDA for expected levels of tritium in water implies reduced sample size.

A companion paper submitted to the J. Radioanal. Nucl. Chem. (\cite{MeasurementsPaper}) evaluates in detail a number of specific cases. A summarized selection of three examples from that analysis is presented here to demonstrate the potential use of a ULB LSC system:

(\textbf{\#1})~A campaign of measurements of tritium backgrounds in natural spring waters was conducted prior to construction of a repository for radioactive waste (\cite{tritiumnaturalsprings}). A total of 124 springs were sampled and each sample was electrolytically enriched in tritium (20$\times$ enrichment). Liquid scintillation counting was employed to measure the resulting tritium concentrations using a LSC system having 20-23\% detection efficiency and a background rate of 1-2~cpm, resulting in an MDA of 0.3~Bq/L. The electrolytic enrichment step required 139~hours/sample and each sample was counted seven times for 30~minutes. Together this is approximately 6~days of total process time per sample. In the case of a low background LSC system with a background rate reduced to 0.01~cpm, the spring water samples could be counted without the electrolytic enrichment step achieving the same MDA in 195~minutes of LSC counting. Including the seven-fold replicate counting, the total process time would remain less than 1~day.
%(\textbf{\#1})~Regulatory limits in the United States state that the tritium activity in a sample must be below 100 Bq/g or become classified as ``Radioisotope'' waste. An LSC system with a background of 6 cpm achieved an MDA of 5.4~$\times$~10$^{-3}$~Bq/g for a 20 gram sample (\cite{tritiumwaste}. The same MDA would be possible using only 3.7 grams of the sample if the background count rate were reduced to 0.1 cpm.
%(\textbf{\#1})~Samples of water and biota taken for environmental monitoring downstream of nuclear power plants were analyzed by noble gas mass spectrometry for their tritium content (\cite{tritiumNPP}). Processing entailed seperating tissue free-water tritium (TFWT) from organically bound tritium (OBT) in reducing the biota samples. Noble gas spectrometry requires the samples are held for 2-3~months after processing to allow $^{3}$H decay into $^{3}$He. Conventional LSC counting would require a count time of 6~hours to obtain an MDA below the upstream, background tritium levels of 1.4~Bq/kg (for 10 mL samples, 1 cpm background, and 21.5\% counting efficiency for tritium). A LSC system with background rates of 0.1~cpm or 0.01~cpm could achieve the same 1.4~Bq/kg MDA with only 120 or 30 minute counting durations, respectively.

(\textbf{\#2})~A method for determination of $^{90}$Sr levels in food samples is comprised of the combustion, dissolution, precipitation, extraction chromatography, ion exchange chromatography, and final precipitation of a 140~g food sample (\cite{SrFood}). This complex process requires two days to convert the food sample into an LSC sample for analysis with a 1220~Quantulus\texttrademark\ low-background LSC. An MDA of 0.1~Bq/kg was achieved with 100~minute count times. Lowering the background to 0.1 or 0.01~cpm would reduce the sample size to 50 or 22~g, respectively, for an equivalent MDA, potentially reducing the scale of the complex sample preparation process.
%(\textbf{\#2})~Sampling of $^{90}$Sr and $^{90}$Y in seawater at stations off the Japanese coast after the Fukushima-Daiichi accident required 50 L of seawater per sample (\cite{SrSeawater}. While maintaining their MDA of 0.4 mBq/L, which is sufficient for meeting background levels in the seawater, and 100 minute count times, the sample size could be reduced to  9 L and 4 L with background count rates of 0.1 and 0.01 cpm, respectively.
%(\textbf{\#2})~A method for determination of $^{90}$Sr levels in food samples is comprised of the combustion, dissolution, precipitation, extraction chromatography, ion exchange chromatography, and final precipitation of a 140 g food sample (\cite{SrFood}). This complex process requires two days to convert the food sample into an LSC sample for analysis with a 1220 Quantulus\texttrademark\ low-background LSC. An MDA of 0.1 Bq/kg was achieved with 100 minute count times. Lowering the background to 0.1 and 0.01 cpm would reduce the sample size to 50 and 22 g, respectively, for an equivalent MDA. The smaller sample sizes may result in faster sample processing and a shorter total time for processing and counting.

(\textbf{\#3})~As part of the $^{235}$U decay chain, $^{227}$Ac is present in both the water column and ocean sediment. The parent $^{231}$Pa is particle-reactive, taking residence in ocean floor sediment, while $^{227}$Ac is non-particle reactive and is released to the water column. Thus $^{227}$Ac concentration in excess of that supported by the $^{235}$U concentration in the water represents a tracer of water transport from the ocean floor on time scales comparable to the 21.8~year half-life of $^{227}$Ac (\cite{AcGeibert2002}). Typical measurements of $^{227}$Ac in this context rely upon chemical isolation of a pure actinium fraction for electroplating followed by a 3-4~month in-growth period after which the more readily measured daughter $^{227}$Th is alpha-counted. The lowest limit of quantification reported in sea water is for an 80-L sample containing $<$10,000 atoms of $^{227}$Ac/L ($<$10~$\mu$Bq/L) (\cite{AcGeibert2008}). Achieving roughly the same level of sensitivity is possible if the purified Ac fraction is counted for approximately 2~days on a LSC system having a background rate of 0.1~cpm, without waiting for daughter in-growth to perform the measurement.

In summary, these three potential applications together begin to demonstrate how an ultra-low background LSC could allow for processing and evaluating samples in significantly shorter \emph{total process time} and/or reduce the sample sizes required in current techniques.

The ultra-low-background instrument suggested above is only achievable by both siting the instrument in a location shielded from cosmic-ray products and incorporating background reduction methods in the system design. For example, a report shows the performance of commercial Quantulus systems, comparatively, in surface and underground locations (\cite{undergrndquantulus2004}). Roughly a factor of three background reduction was obtained in the Quantulus system operated underground. The report also studied the impact of scintillation vial volume (1-7~mL) on detection sensitivity for $^{14}$C demonstrating 0.1-0.3~cpm background rates are achievable in the underground location. The 3800 meters-water-equivalent (mwe) overburden shielding largely eliminates the background rate due to cosmic-ray products and thus implies the limiting backgrounds for the system are the naturally occurring radioactivity in the laboratory and/or the instrument's components. The present article presents a design analysis for a LSC system targeting a background rate in the range of 10-100 counts per day (cpd), equivalent to 0.07-0.007~cpm.

% Please send JLO the article: \bibitem[Kalin and Long, (1989)]{undergrndquantulus1989} Kalin, R.M. and A. Long, 1989, Radiocarbon Dating with the Quantulus in an Underground Counting Laboratory: Performance and Background Sources, Radiocarbon. 31 3. pp. 359-367

\section{ULB LSC shield design}
\label{ShieldDesign}

The ULB LSC shield design is based on a number of general principles, experience with similar shields already present in the Pacific Northwest National Laboratory (PNNL) shallow underground laboratory, and a few initial parametric simulation studies. Each of these three types of consideration is presented in-turn within this section. The key innovation for the ULB LSC system is the implementation of low background shielding best practices around a novel approach to the scintillation light collection (\cite{OpticsPaper}). A hollow light guide cut out of the copper interior shielding and coated with a specularly reflective material will guide the photons emitted from the liquid scintillator around a 90-degree curve. The curved shape minimizes interaction of the intrinsic radioactivity in the PMTs with the scintillation volume while maintaining light collection efficiency for decay events in the sample volume. This section presents the basis for the initial shield development and concludes with a detailed description of the shield design that leads to a more refined background estimation.

\subsection{Low background shielding principles}
\label{subsec:principles}
The nature of low-background measurements necessitates the shielding design follow a set of guiding principles. The starting point for shielding gamma rays in any system is using a high-density material, often lead, to block external gamma rays from the detector. The decay of $^{210}$Pb and daughters, however, induces a bremsstrahlung background within the lead shield. In low-background systems, high purity Oxygen-Free High-Conductivity (OFHC) copper is often used as a liner inside the lead shielding to attenuate the bremsstrahlung radiation. The amount of OFHC copper needed to reduce the induced bremsstrahlung background is reduced by inserting a layer of low-background lead into the shield design. A combination of these shielding materials will create an effective shield against gamma backgrounds.

Shielding against cosmic-ray by-products (protons, neutrons, and muons) is best accomplished by locating the system underground. The first few meters of overburden will screen-out protons, but tens of meters of overburden are required to reduce the cosmic-ray-induced neutron flux to a level at or below the level of neutrons emitted by the surrounding rock. The environmental neutrons produced through $(\alpha,n)$ reactions of U and Th chains in the surrounding rock and concrete, as well as those from the spallation of cosmic-ray by-products, are reduced using a combination of a neutron moderator (water or polyethylene) and a neutron absorbent material (boron or cadmium). The muon background cannot be removed by introducing absorbing materials or moderators. Instead, scintillation panels are used as an active veto to tag counts coincident with muon events. Ideally these scintillation panels provide complete 4$\pi$ coverage and are sufficiently thick to provide energy discrimination between muons passing through the plastic and environmental 2615~keV gamma-rays. This is typically accomplished with 2$^{\prime\prime}$-thick plastic scintillator panels.

\subsection{Results from other shallow underground shields}
\label{subsec:formerresults}

Given the above general low background design principles, prior designs of low background systems were considered as points of reference for the ULB LSC development. There are two existing low background shields in the PNNL shallow underground lab that have had sufficient data collection to draw some inferences on design choices.

An array of fourteen high-purity germanium (HPGe) detectors (CASCADES) (\cite{CASCADES}) is used for sample counting and is housed within a necessarily large lead shield. It has been observed there are neutron scattering features in the HPGe spectra collected from the instrument. It is believed the amount of lead present in the shield is acting as a spallation target for muons, producing the high-energy neutrons that are seen in the HPGe detector array. Thus, if it is possible to maintain a compact shield with only the minimum required lead thickness, it is possible to limit the muon-produced neutron background (\cite{HPGeShield}).

The activation of copper has been a deterrent to using an inner copper liner for bremsstrahlung shielding in some low-background systems (\cite{cebrianRef}). While these concerns may be valid for surface level systems, simulations with high energy spallation neutrons have yielded the copper activation levels shown in Table~\ref{tab:CuActLevels} for a depth of 75~mwe (\cite{copperactivation}). These numbers show that the background rate from copper activation is small relative to the other contributors. In addition, placing a 2.5-cm-thick OFHC copper shield inside the CASCADES system lowered the overall background level observed in the HPGe crystals. It was concluded that the addition of the copper shield helped to mitigate the bremsstrahlung photon background from the lead (\cite{shallowlab}) without adding an additional background component. The low activation levels and clear shielding benefits of using an inner copper layer promoted the incorporation of an inner copper shield into the ULB LSC. 

\begin{table}[ht!]
\begin{center}
\caption{\label{tab:CuActLevels}
Simulated copper activation levels from high energy spallation neutrons for a 75~mwe underground facility.}
\begin{tabular}{ cc }
\multicolumn{2}{c}{} \\ \hline \hline
Isotope & Activation Level \\
 & (atoms/kg/year) \\ \hline
$^{46}$Sc & 2.32 \\
$^{51}$Cr & 15.32 \\
$^{52}$Mn & 2.69 \\
$^{54}$Mn & 14.10 \\
$^{59}$Fe & 2.64 \\
$^{56}$Co & 4.84 \\
$^{57}$Co & 18.92 \\
$^{58}$Co & 45.33 \\
$^{60}$Co & 29.73  \\ \hline \hline
\multicolumn{2}{c}{} \\
\end{tabular}
\end{center}
\end{table}

An array of twelve gas proportional counters (\cite{ULBPC}) has provided two pieces of relevant information: First, by necessity of access, one of the six sides of the shield is not covered by a plastic scintillator panel. It is believed this accounts for a residual several dozen counts per day in each proportional counter operated in this shield. This confirms the principle of when feasible, ensure a nearly full coverage by plastic scintillator panels. Second, the proportional counter array was used to make a measurement of the neutron flux within the shield by filling a proportional counter with $^{3}$He. This resulted in a thermal neutron flux measurement of $(2.4 \pm 0.4) \times 10^{-6}$~neutrons/cm$^{2}$/second (\cite{KosHe3}). The proportional counter shield is \emph{very} similar to the shield plan for the ULB LSC system and thus the neutron flux measurement provides a method for estimating the expected thermal neutron-induced background rate. Since the proportional counter array employs 2$^{\prime\prime}$-thick, 30\% borated polyethylene sheets to moderate and absorb neutrons, the ULB LSC shield should use at minimum these same features to reach a neutron flux similar to that measured in the proportional counter shield.

\subsection{Parametric simulation studies}
\label{subsec:parametricsims}

To inform the thicknesses of the graded shield layers, simple GEANT4 simulations were performed to assess the backgrounds produced by the room gamma-ray background, $^{210}$Pb activity in the lead shield, and a low background photomultiplier tube. GEANT4 is an object-oriented, C++-based radiation transport simulation toolkit (\cite{GEANT-NIM,GEANT-IEEE}). Optical, electromagnetic, and hadronic processes as well as radioactive decay are incorporated into the physics simulation packages, and can be implemented within complex model geometries. For the radiation transport simulation of the ULB LSC system, the GEANT4 radioactive decay and low energy electromagnetic processes are employed to create sources of background within the geometric model of the ULB LSC shield, described in Section~\ref{subsec:SimGeo}. These initial simulation studies were parametric in nature, determining the anticipated background rates for a series of varying thickness of shielding components. These simulations provided the intuition to begin the shield design with a notion of the rough order of magnitude backgrounds expected for the shield design concept.

\begin{figure}[ht!]
\begin{center}
\includegraphics[width=\columnwidth]{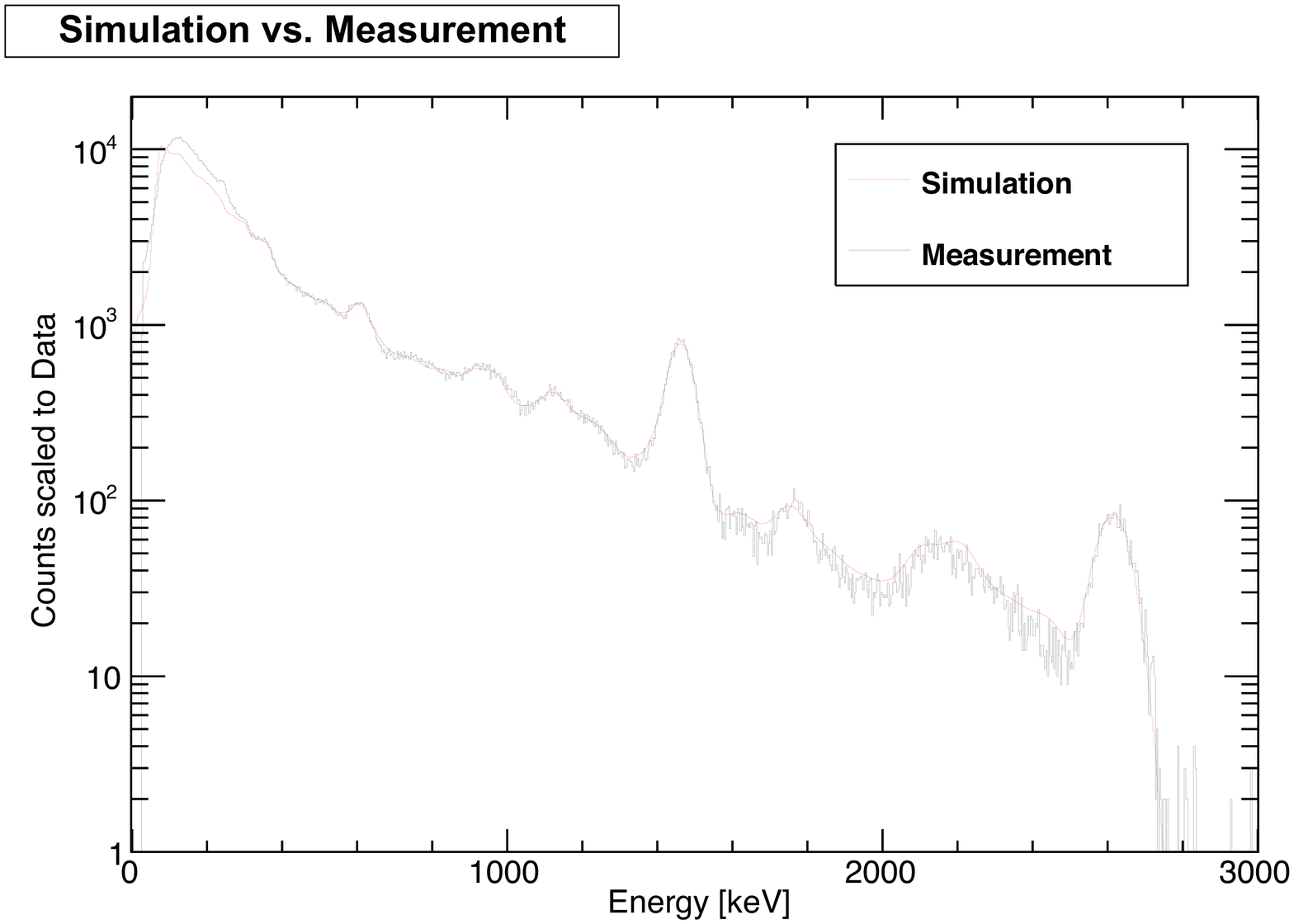}
\caption{\label{fig:backgroundcomp} Comparison of simulated and measured room background for 3$^{\prime\prime}$~$\times$~3$^{\prime\prime}$ NaI detector.}
\end{center}
\end{figure}

To benchmark the background gamma-ray flux in the underground laboratory, a measurement was made with a 3$^{\prime\prime}$~$\times$~3$^{\prime\prime}$ NaI detector. A simulated U/Th/K-based background spectrum was then scaled to match the measurement. The U/Th/K contributions were independently scaled in the simulation to fit the collected data for the 3$^{\prime\prime}$~$\times$~3$^{\prime\prime}$ NaI detector, the collected data and scaled-fit are shown in Figure~\ref{fig:backgroundcomp}. The fit matches the data over the range from 300-2700~keV. This is believed to provide a reasonable representation of the external gamma ray flux relevant to a lead shield. This scaling was used to determine the background rate in 24~mL of liquid scintillation cocktail through varying thicknesses of shielding materials (lead with an inner liner of copper). Table~\ref{tab:parametricsims} shows the results from this measurement and scaled-simulation effort showing the amount of shielding required to obtain a given total background count rate from external gamma rays. A natural extension to this simulation was to include the impact of $^{210}$Pb activity in the lead bricks. A 60~Bq/kg background from $^{210}$Pb (and daughters) as a 0.356~particles/s/m$^2$/(Bq/kg) surface source (\cite{LeadSource}) was assumed for the lead bricks used in the shield design. Table~\ref{tab:parametricsims} also shows the inner copper layer thickness required to screen out the induced bremsstrahlung background from $^{210}$Pb and daughter decays.
 
A separate, simple simulation was done to assess the background associated with the U/Th/K content of a low background photomultiplier tube. For the Electron Tubes 9266XXXB model photomultiplier (\cite{electrontubes9266B,background9266B}) an un-shielded tube is expected to contribute roughly 700 counts per day, thus pointing to the need to use light guides to ``hide'' the radioactivity of the photomultiplier tube behind some lead shielding.

\begin{table}[ht!]
\begin{center}
\caption{\label{tab:parametricsims}
Estimated background count rates in counts per day (cpd) for the parametric simulation studies of shielding against three of the most important backgrounds. A dash (``--'') indicates no simulation was performed with this shielding thickness. A 24-mL scintillation cocktail volume was used as the target volume in these simulations.}
\begin{tabular}{ |c|c|c|cc| }
\multicolumn{5}{c}{} \\ \hline
Background\vphantom{\rule{0cm}{2.5ex}}   & $^{210}$Pb & External & \multicolumn{2}{c|}{PMT} \\
source      & Brem.          & $\gamma$-rays & \multicolumn{2}{c|}{(9266B)} \\ \hline \hline
Shielding\vphantom{\rule{0cm}{2.5ex}}       & Cu               & Pb with  &  Cu & Pb \\
material    &                    & 8~cm Cu liner    &                    &      \\ \hline \hline
Shielding\vphantom{\rule{0cm}{2.5ex}}       & \multicolumn{4}{c|}{Parametric background rate} \\
material                                                      & \multicolumn{4}{c|}{estimates in counts per day} \\
thickness                                                    & \multicolumn{4}{c|}{(cpd)} \\ \hline
0 cm\vphantom{\rule{0cm}{2.5ex}}  & 3,525 & 39,938 & 698 & 698\\
1 cm  & 474 & --       & 201 & 79 \\
2 cm  & 275 & --       & 111 & 27 \\
3 cm  & 156 & --       & 58  & 12 \\
4 cm  & 86   & 2,327 & 31 & 6 \\
6 cm  & 46   & --      & 12 & 2 \\
8 cm  & 26   & 226    & 4   & -- \\
10 cm & 14  & --      & -- & -- \\
12 cm & 8    & 37     & -- & -- \\
14 cm & --  & 13      & -- & -- \\
16 cm & --  & 5        & -- & -- \\ \hline
\multicolumn{5}{c}{} \\
\end{tabular}
\end{center}
\end{table}

\subsection{Low background LSC shield design concept}
\label{subsec:Shield}

\begin{figure}[ht!]
\begin{center}
\includegraphics[width=\columnwidth]{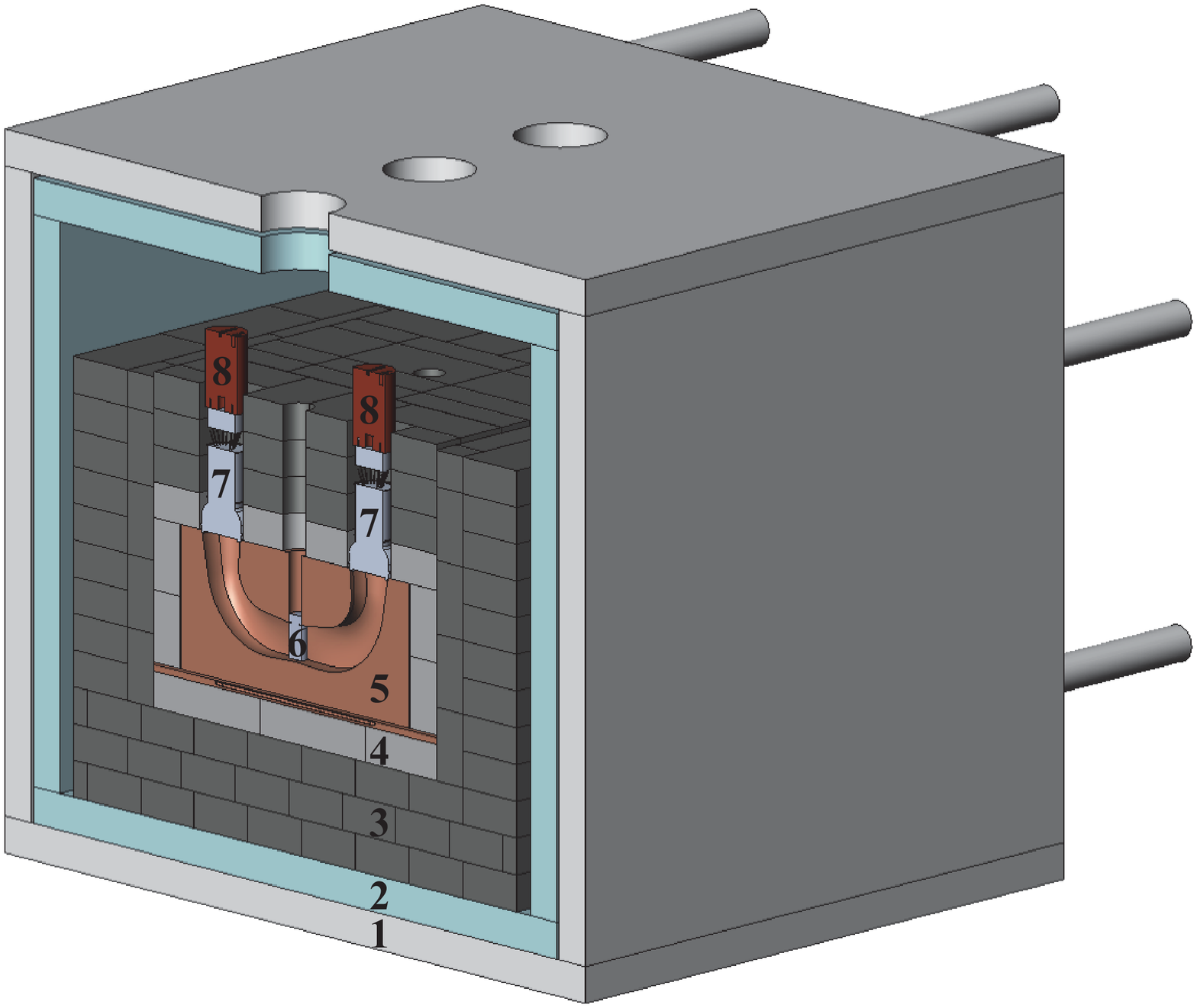}
\caption{\label{fig:innershield} Cut away view of the shield components. (1) Outer layer of plastic scintillator veto panels (grey). (2) Borated polyethylene (teal). (3) Lead shielding (dark grey). (4) Low background lead shielding (grey). (5) Hollow copper light guide (orange). (6) Liquid scintillation vial (grey). (7) Photomultiplier tubes (PMTs) (light grey). (8) PMT bases (red). See text for additional description.}
\end{center}
\end{figure}

The concept for the low background LSC system is shown in Figure~\ref{fig:innershield}. The graded shield concept includes an exterior lead shield with a lower background lead liner as well as a high-purity inner copper measurement chamber. The inner copper also forms light guides that transport the scintillation photons from the vial to two photomultiplier tubes. The entire lead shield is surrounded by borated polyethylene and plastic scintillator.  Not shown in Figure~\ref{fig:innershield} is the radon exclusion box and sample introduction mechanism. The design includes three counting positions: left, center, and right chambers. This design is further informed by the investigations of the following sections of this article.

\section{Background estimates}
\label{sec:BackgroundEstimates}

A more complete simulation effort was undertaken during the design period to evaluate the anticipated backgrounds of the shield design presented in Sec.~\ref{subsec:Shield}. During this design phase of the project a medium-fidelity shield model within GEANT4 was developed to rapidly provide feedback to the design work. The GEANT4 model of the shield represents all the major features of the shield but is deliberately limited to construction from rectilinear and right cylindrical shapes to speed the development of simulation results to inform the progression of the shield design.

\subsection{Simulation geometry}
\label{subsec:SimGeo}
The geometry used a set of nested Logical Volumes to create the different layers of shielding. The innermost copper shield was nested within layers of low background lead, surrounded by the bulk lead shielding. The shield design includes three counting chambers. Within the simulation, the left and right chambers are modeled assuming a PMT-based light detection system. The simulation model treats the central measurement chamber as a fully shielded volume containing a liquid scintillation vial. This fully shielded vial provides a best-case scenario for comparison within the radiation transport simulation and is used to evaluate the loss in shielding effectiveness due to the penetrations required for the PMTs in the left and right measurement chambers.\footnote{Incidentally, deliberately omitting a model of light detection instrumentation within the central measurement chamber is appropriate. The ULB LSC development plan is intentionally reserving the central chamber for future development and improvements once experimental evaluation of the left and right measurement chambers' performance is obtained.} In each of the PMT-based systems, two low background lead bricks are inserted into the copper shielding between the vial and either PMTs to place more attenuating lead shielding directly between the vial and PMTs. The PMT cavities were cut out of the shielding volumes to avoid overlap with and undue complexity in the shielding volumes. The complexity and number of volumes used were minimized to reduce the run time for the simulations. The dimensions for the simulation geometry are given in Table~\ref{tab:dimensions}. The material characteristics are shown in Table~\ref{tab:materials}. Pure copper, lead, and aluminum were used for the copper shield, lead shields, and PMT tubes and bases respectively. The density of the aluminum was adjusted to accurately portray the amount of aluminum in the PMT volume (labeled LightAl). Silicon dioxide was used for the PMT glass, and air for the world volume and cavities in the shields, PMTs, and vials. The vials were simulated with a polyethylene terephthalate (PET) wall and filled with a cocktail liquid (LSC2) composed of C, H, N, and O. The LSC2 material definition describes an approximation of a 100 g/L of common liquid scintillator containing 2,5-diphenyloxazole (PPO) in a toluene solution. This material definition is in general an adequate representation of liquid scintillation cocktails for the purpose of tracking energy depositions within materials using radiation transport simulation software. To ensure a common metric, the results from each simulation were reported in terms of ``counts per day'' (cpd), here defined to mean when any energy is deposited in the scintillation cocktail volume during the transport of an initially thrown starting radiation emission.

\begin{figure}[ht!]
\begin{center}
\includegraphics[width=\columnwidth]{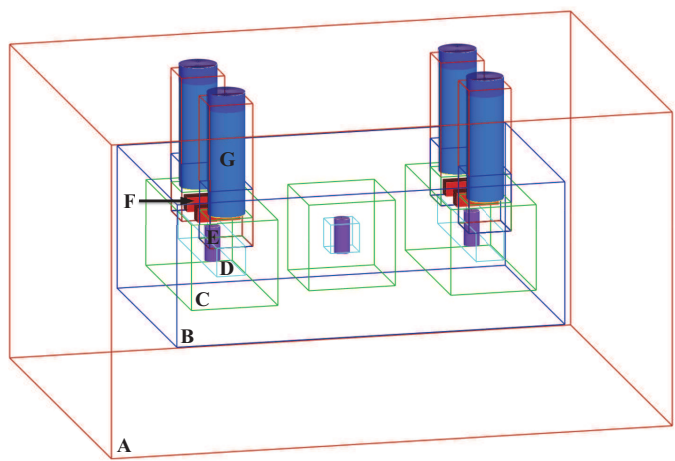}
\caption{\label{fig:innershieldGEANT} Simulation view of the medium fidelity GEANT4 model of the low background shield. The shield simulation model is composed of the following volumes: (A) The outer most rectilinear volume (red line) is bulk lead. (B) The interior rectilinear volume (purple line) is low background lead. (C) The three rectilinear volumes (green line) are copper shielding. (D) Rectilinear volumes (blue line) represent the hollow portion of the light guide system. (E) Solid right cylinders (purple solid) represent the scintillation vials and are used to record energy depositions with the radiation transport simulation. (F) Four rectilinear volumes (red solid) are low background lead bricks placed directly between the photomultiplier tubes (PMTs) and liquid scintillation vials. (G) The four solid right cylinders (blue solid) are the PMTs instrumenting the left- and right-most measurement chambers. The central measurement chamber is not instrumented and serves as a comparison as a fully shielded volume.}
\end{center}
\end{figure}

\begin{table}[ht!]
\begin{center} 
\caption{\label{tab:dimensions}
The dimensions of the outer extent of the simulation volumes used to represent a medium fidelity model of the low background shield in GEANT4. See text and Figure~\ref{fig:innershieldGEANT} for additional explanation.}
\begin{tabular}{l c c c }
 & & & \\ \hline \hline
Model Components (\#) & Width & Height & Depth \\ 
(nested cuboids) & (cm)  & (cm)  & (cm) \\ \hline
Bulk lead (1) & 99.06 & 55.88 & 73.66 \\
Low background lead (1) & 68.58 & 25.4 & 43.18 \\
Left \& right copper (2) & 15.24 & 15.24 & 33.02 \\
Left \& right chamber (2) & 5.08 & 5.08 & 27.94 \\
Central copper (1) & 15.24 & 15.24 & 15.24 \\
Central chamber (1) & 5.08 & 5.08 & 5.08 \\
PMT ports  (4) & 7.62 & 25.40 & 7.62 \\
Line-of-sight lead (4) & 7.62 & 2.54 & 2.54 \\ \hline \hline
 & & & \\ \hline \hline
Vials \& PMTs (\#) & Diameter & Height &  \\
(right cylinders) & (cm) & (cm)  & \\ \hline
PMTs (4) & 6.6 & 21.8 & \\ 
Vials (3) & 2.70 & 5.0 & \\ \hline \hline
 & & & \\
 & & & \\
\end{tabular}
\end{center}
\end{table}

\begin{table}[ht!]
\begin{center}
\caption{\label{tab:materials} Material compositions in GEANT4 simulations.}
\begin{tabular}{l r r c}
& & &  \\ \hline \hline
Material & \multicolumn{2}{c}{Elemental} & Density \\
             & \multicolumn{2}{c}{Fraction} & (g/cm$^3$) \\ \hline
Air         & N   & 78.44 \% & $1.205\times10^{-3}$ \\
             & O   & 21.08 \% & \\
             & Ar  & 0.47 \%  & \\
             & C   & 0.02 \%  & \\  \hline
Copper  & Cu & 100 \%  & 8.96 \\  \hline
Glass     & Si  & 33.3 \% & 2.6 \\
             & O  & 66.7 \% & \\ \hline
LightAl  & Al  & 100 \%  & 0.445 \\ \hline
LSC2     & C   & 47.08 \% & 0.96 \\
             & H  & 52.03 \% & \\
             & N  & 0.59  \% & \\
             & O  & 0.30 \% & \\  \hline
Lead      & Pb & 100 \% & 11.34 \\ \hline
PET       & C   & 33.3 \% & 1.32 \\ 
             & H  & 33.3 \% & \\
             & O  & 33.3 \% & \\ \hline \hline
\end{tabular}
\end{center}
\end{table}

\subsection{External gamma-ray backgrounds} %% LSC External Gamma.v0.docx; LSC External Gamma.v0.JLO.docx
\label{subsec:ExtGamma}
External gamma rays are principally due to U- and Th-chain radioisotopes and radioactive $^{40}$K present in the rock and concrete surrounding the shallow underground laboratory. The background source term is a gamma-ray spectrum developed to match a background spectrum previously acquired in the underground laboratory, as described above in Sec.~\ref{subsec:parametricsims}. The external gamma source was modeled by placing the source term on the inner edge of a 150~cm $\times$ 150~cm $\times$ 150~cm cubic world volume surrounding the shield. The initial geometry had cavities for the PMTs, but no covering above the PMT ports, yielding a background contribution of $321 \pm 27$~cpd for each of the vials in the left and right chambers. The central vial did not register any events, implying further study of placing lead shielding above the PMT ports would be fruitful. Of the 321~cpd, 97.6\% of the events had originated directly above the shielding, and the majority of the remainder came in at an angle from above the shielding. By adding lead end caps over the PMT ports, this count rate is reduced. Table~\ref{tab:contributors} shows the decrease in the contribution to background count rate as the thickness of lead above the PMTs increases. The final shield design will thus include as much additional lead shielding above the PMTs ports as is feasible.

\begin{table}[ht!]
\begin{center}
\caption{\label{tab:contributors} Background count rate contributions of external gamma rays as varied thicknesses of lead end caps are placed above the PMT ports in the shield.}
\begin{tabular}{c r@{ $\pm$ }l}
                & \multicolumn{2}{c}{} \\ \hline \hline
Thickness & \multicolumn{2}{c}{Background} \\
(inches)    &  \multicolumn{2}{c}{(cpd)} \\ \hline
0.0$^{\prime\prime}$ & \phantom{--}321 & 27 \\
0.5$^{\prime\prime}$ & 94 & 10 \\
1.0$^{\prime\prime}$ & 34 & 6 \\
1.5$^{\prime\prime}$ & 25 & 5 \\
4.0$^{\prime\prime}$ & 2 & 1 \\\hline \hline
\end{tabular}
\end{center}
\end{table}

\subsection{Radioactivity in the lead shield} %%LSC Pb Shielding Background.v0.docx
\label{subsec:Pb210}
While the lead shield reduces the external background observed in the vials, the background radiating from the lead itself, namely resulting from $^{210}$Pb and $^{210}$Bi decays, must be characterized and minimized. The approach taken in the design for minimizing this background is a graded system of an outer lead shield, middle shield of low background lead, and inner shield of copper to attenuate the bremsstrahlung radiation escaping the lead. The emissions from the lead have been previously characterized in literature (\cite{LeadSource}). In this paper, the normalized surface source is 0.356~particles/s/m$^2$/(Bq/kg), and the specific activities assumed for normal and low background lead are 60~Bq/kg and 3~Bq/kg, respectively. An lead assay campaign that is intended for future publication located available lots of lead having 75~Bq/kg $^{210}$Pb for the outer lead. The inner layer of 3~Bq/kg $^{210}$Pb is available for purchase, though at a premium price.

The surface sources were placed inside the outer and middle shields, on the extra lead brick shields between the PMTs and vials, and around the holes for the PMTs in the middle and outer shields. The emission surface for the outer lead was placed immediately around the inner lead shield and the surfaces of the PMT ports. The emission surface for the low background lead was placed immediately around the copper volumes and on the surface of the four line-of-sight lead bricks.

Where possible, geometric symmetry was relied upon to simplify the number of simulations run. For instance, a single run was completed for the contributions of one of the lead blocks inserted between the vial and PMTs, and those results were doubled in order to account for the two lead bricks in a light guide system (one between each of the PMTs and the vial). The lead end caps above the PMT ports described previously (Sec.~\ref{subsec:ExtGamma}) were included in the geometry, but not modeled as a surface source. Contributions from the inside of the outer and middle shields in the PMT holes were included, and it is assumed that these surface contributions will compensate for the excluded lead end cap surfaces. After an equivalent of 10~days of exposure, the normalized background count rates were $3.4 \pm 0.6$~cpd for the vials in the left \& right chambers and $0.6 \pm 0.2$~cpd for the central vial. The elevated count rate in the outer vials is primarily due to the surface sources in the PMT ports. This number is expected to decrease if low background lead is used to line the PMT ports, which has become the resulting construction plan.

\subsection{Radioactivity in the OFHC copper} %%LSC Cu Background.v0.docx; LSC Light Guide Background.v0.docx
\label{subsec:OFHC}
Commercially-available high purity OFHC copper (alloy C10100) is used for the inner shield and reflective hollow light guide. This OFHC copper contains trace levels of naturally occurring radioactive material (specifically U/Th/K). The copper shields around the left, right, and central blocks were simulated, conservatively, as containing order of magnitude concentrations of U/Th/K at the 10~ppt levels for $^{238}$U/$^{232}$Th and 100~ppb for $^{40}$K. Decay chains were included and assumed to be in secular equilibrium. These values for U/Th/K concentrations are based on rough expectations (upper limit concentrations) for high purity OFHC copper available from manufacturers (\cite{exoRef}). Plans for fabrication of the LSC system include U/Th assay of the procured OFHC copper to determine the actual U/Th concentration in the LSC copper shielding. The simulation also includes a level of $^{60}$Co due to the activation of the OFHC copper from cosmic rays. A sea level cosmogenic production rate of 97.4~$^{60}$Co/kg/day (\cite{cebrianRef}) is used to determine the anticipated activity in the copper shielding assuming 6~months of cosmic ray exposure. The left and right copper shields contain the bent light guide systems while the central copper shield contains only the sample vial. The simulation results show a background count rate of $6.6 \pm 0.26$~cpd for the left or right vials and $7.85 \pm 0.28$~cpd for the central sample. The background is due entirely to the surrounding copper shielding. No cross-talk was observed between the copper shielding of the outer vials and the central sample vial, or between the central copper shielding and the outer vials.
% values changed from 1.2$\pm$0.1 cpd and 1.3$\pm$0.1 cpd, respectively, from calculations based on LSC Cu Background.v0.docx. Calcs in ShieldPaperCounts.xls.

\subsection{Radioactivity in the reflective coating}
\label{subsec:Coating}
Background contributions from the reflective coating, a series of Ta$_{2}$O$_{5}$ and SiO$_{2}$ layers, in the copper light guide were simulated. The source terms were estimated from the known impurities in the precursor to Ta$_{2}$O$_{5}$, tantalum (V) ethoxide (Ta(C$_{2}$H$_{5}$O)$_{5}$). A previous analysis of tantalum (V) ethoxide found the U/Th content to be \textless0.04~ng/g (0.047~pg/cm$^2$) in the material (\cite{KozonoRef}). The vendor's chemical analysis yielded \textless1~ppm (1.169~ng/cm$^2$) of natural potassium (\cite{TantRef}). The impurity levels in the SiO$_{2}$ precursor, Di-t-butoxydiacetoxysilane, were assumed to be equal to the those of the tantalum (V) ethoxide, and the coating materials were assumed to have the same impurities as their precursors.  Using 0.012~pg/cm$^2$ of $^{238}$U and $^{232}$Th (with decay chains in secular equilibrium) and 0.3~ng/cm$^2$ of natural K, the coating was simulated as a surface source along the hollow horizontal channel, the vertical channels leading to the PMTs, and the outer surface of the copper block. These surface coatings reflect the low pressure chemical vapor deposition (LPCVD) process expected to be used for the reflective coating application that will subsequently coat the entire copper block. The number of events simulated was equivalent to 1000~days of decay for the surface sources.

The simulation results showed a background rate of $0.008 \pm 0.001$~cpd. Of this background rate, 99.6\% originated from the interior of the light guide channel; the outer coated surface of the copper shield contributes very little to the total count rate. Furthermore the $^{40}$K decays dominated contributing $0.0079 \pm 0.001$~cpd to the background rate. The equivalence assumption of the U/Th/K concentration levels in the reflective coating materials is not the most accurate basis for simulation sources. Even so, the results can be off by a factor of 100 before this background becomes a legitimate concern. Additionally, coating the copper via the LPCVD process is expected to reduce the impurity concentrations in the reflective coating. The impurity reduction is unaccounted for in this background estimation due to a lack of data on changes in impurity concentrations in the LPCVD process. The simulation estimates demonstrate the small background contribution from the reflective coating with regard to the other background sources. Laser ablation inductively coupled plasma mass spectrometry (LA-ICP-MS) or acid dissolution ICP-MS analysis of a test coating is planned to determine the actual impurity levels in the reflective coating application.

\subsection{Radioactivity in the PMTs} %%LSC PMT Background.v0.docx
\label{subsec:PMTs}
Natural radioactivity in the PMT was simulated as $^{40}$K, $^{60}$Co, and the $^{238}$U and $^{232}$Th decay chains (in secular equilibrium) concentrated in the PMT glass. The glass was modeled with a 64~mm diameter and 2~mm thickness, similar to the dimensions of the Hamamatsu R11065 and R11410 (\cite{HamamatsuR11065}, \cite{HamamatsuR11410}). The concentrations used for the radioisotopes in the PMTs were 8.3~mBq/PMT of $^{40}$K, 2~mBq/PMT of $^{60}$Co, 0.4~mBq/PMT of $^{238}$U, and 0.3~mBq/PMT of $^{232}$Th which were taken from reports from the high energy physics community (\cite{LUXPMTs}).  Evaluation of the PMTs shows the assumption of uniform distribution of the radioisotopes in the PMT glass envelope to be conservative as most of the natural radiation will be contained in the ceramic section toward the back end of the PMT. Simulations were run for a single PMT with 1000~days of exposure. The results were normalized with respect to time and multiplied by 2 to account for radiation from both PMTs in each measurement system. The background count rate due to activity in the two PMTs is $1.6 \pm 0.1$~cpd. The primary contributor was $^{60}$Co. From the simulations, it was also observed that there is no cross-talk between the PMTs on the left and right chamber-wings and the central sample vial.

\subsection{Background from the vial} %%e20140602-VialBackgroundEstimate.pdf
\label{subsec:Vial}
The sample vials for the liquid scintillation counting will inherently contain some U/Th/K. Using the mass of the vial and specific activity of the U/Th/K, a simple calculation can estimate the background from the sample vial assuming part per trillion (ppt) levels of U and Th. The following assumptions are applied for the calculation: the mass of the vial is 10~g and the activity of $^{238}$U is 12.5~$\mu$Bq/kg, with 10 daughters in secular equilibrium (hence a total of $\sim$125~$\mu$Bq/kg). Equation~\ref{eq:Vial} shows the background rate from the $^{238}$U decay chain in vial is 1.25~$\mu$Bq, or 0.1 decays per day:
\begin{eqnarray}\label{eq:Vial}
\lefteqn{\textrm{Activity} ~ \bigg(\frac{\textrm{decays}}{\textrm{day}}\bigg)}\nonumber \\
 &=& \begin{array}{c}\textrm{Specific}\\\textrm{Activity}\end{array} \bigg(\frac{\mu\textrm{Bq}}{\textrm{kg}}\bigg) ~ \times ~ \textrm{Mass} ~ \bigg(\textrm{kg}\bigg) ~ \times ~ 86400 ~ \bigg(\frac{\textrm{s}}{\textrm{day}}\bigg)\nonumber \\
 &=& 125 ~ \bigg(\frac{\mu\textrm{Bq}}{\textrm{kg}}\bigg) ~ \times ~ 0.01 ~ \bigg(\textrm{kg}\bigg) ~ \times ~ 86400 ~ \bigg(\frac{\textrm{s}}{\textrm{day}}\bigg)\nonumber \\
 &=& 0.1 ~ \bigg(\frac{\textrm{decays}}{\textrm{day}}\bigg)
\end{eqnarray}
The $^{232}$Th decay chain and $^{40}$K will add about half of the activity of the $^{238}$U decay chain, leading to a total count rate of $\sim$0.15 decays per day. Due to the less than 100\% detection efficiency for decays in the vial walls, the counts per day will be further reduced from the total of 0.15 decays per day, thus the background is sufficiently small in comparison to other backgrounds studied above.

\subsection{Cross talk between scintillation vials} %%LSC Cross Talk.v0.docx
\label{subsec:CrossTalk}
When the system is in use, vials containing radioactive samples may be present in all of the counting positions. The possibility of cross-talk between the liquid scintillators was addressed by simulating the outer vial response to an isotropic gamma source in the central vial. The central vial contained a 2615~keV gamma source, emitting 10~$\gamma$-rays per minute, simulated for 1000~days of exposure. The count rate observed in the outer vials for this source is $0.0295 \pm 0.0038$~cpd. This background rate increases to 3~cpd if the shield materials between the sample chambers is replaced with vacuum. These results for a high energy gamma emitter represent a conservative estimate for the level of possible cross-talk between sample chambers.

\subsection{Neutron background estimation}
\label{subset:NeutronEst}
The neutron background flux measured in the underground proportional counter system at PNNL, described in Sec.~\ref{subsec:formerresults}, was used to calculate the estimated neutron background observed for the liquid scintillation cocktail. The cross-sections and reaction rates were calculated using 20~mL of the liquid scintillation cocktail defined for the GEANT4 simulations previously discussed, consisting of O, N, C, and H. The following isotopes are included in the cocktail description: $^{12}$C, $^{13}$C, $^{1}$H, $^{2}$H, $^{16}$O, $^{17}$O, $^{18}$O, $^{14}$N, $^{15}$N. The total macroscopic neutron capture cross-section for 20~mL of the scintillation cocktail is calculated as equal to 0.317~cm$^2$. Taking the neutron flux multiplied by the number of atoms and cross-section of the represented isotopes in a 20~mL cocktail volume, yields a reaction rate of $(7.60 \pm 1.27) \times 10^{-7}$ neutron captures/second. Over a period of one day, the resulting background is estimated as equal to $0.066 \pm 0.011$~cpd.

Thermal neutron capture on $^{63}$Cu produces $^{64}$Cu. Positron emission is one of the decay pathways for $^{64}$Cu, which yields two 511~keV gamma-rays per decay. In the 76.7 kg of copper per light guide, $^{63}$Cu accounts for 69\% (53~kg). The thermal neutron capture cross-section is 4.4~barns resulting in a total thermal neutron capture rate of 462 captures per day in the light guide. Of the 462 activated copper nuclei, 19\% decay by positron emission. A simulation of $^{64}$Cu distributed in the copper volume yielded an efficiency of 0.009\% for counts in the vial from $^{64}$Cu decays. Taking into account the branching ratio, two gammas per decay, and efficiency for resulting counts in the vial, the final background count rate from $^{64}$Cu is 0.016~cpd.

Inelastic scattering on lead nuclei is another potential background from neutrons in the underground environment. Inelastic scattering of neutrons on lead produces a dominant 803~keV gamma-ray that may reach the liquid scintillation cocktail producing unwanted scintillation events. To estimate this background level, an initial assumption was made that only the inner 2~inches of lead would contribute to the background rate. This amounts to a total lead target mass of 212 kg. To estimate the efficiency for generating a background event, two (extrema) situations were considered: (1) All 803~keV gamma-rays are emitted isotropically from the inner surface of the inner-most 2~inches of lead and (2) All 803~keV gamma-rays are emitted isotropically from the outer surface of the inner-most 2~inches of lead. For each thrown 803~keV gamma-ray, the efficiency to produce an energy deposition in the LSC cocktail was $2.5 \times 10^{-4}$ and $1.5 \times 10^{-6}$ the considered emission geometries, respectively. Since 2~inches of lead reduces the efficiency to produce a background event by a factor of 17, but the outer-most lead is only a factor of 7 more in mass, this justifies the initial assumption that only the inner 2~inches of lead are important for the inelastic neutron scattering background contribution. Conservatively, using the highest efficiency results in 0.027~cpd background rate contribution, a small contribution to the overall predicted background rate. These estimations show that the neutron background in the ULB LSC is 0.109 cpd, and likely not one of the larger contributors.

\subsection{Background summary}
The results reported in Table~\ref{tab:backgroundsum} show the background count rate contributions from the sources inherent to the system described in the previous sections. The total background count rate for the outer liquid scintillators is $14 \pm 1.2$~cpd. These simulations have been informative and integral to the development of the shield design. These medium fidelity simulation results have led to reconsideration of the design of the PMT ports, the choice and location of low background lead in the PMT ports, and the need to assay and understand the lead and copper impurity levels to know their final contribution to the system background. Further background characterizations will include the background contributions from muon-induced neutrons, fission neutrons, and neutron production due to alpha captures. Also not captured in this study are \emph{any} background rates associated with systematic effects other than radiation, such as PMT dark current rates and chemiluminescence in the vial, to name just two.

\begin{table}[ht!]
\begin{center}
\caption{\label{tab:backgroundsum} Background contributors to the sample vials in the left or right measurement chambers due to ubiquitous radioactive isotopes in the environment and instrument construction materials. The background rate is estimated in counts per day (cpd).}
\begin{tabular}{l c r}
& &   \\ \hline \hline
Background source  & Rate & Fraction \\
(contributing isotopes) & (cpd) & (\%)  \\ \hline
External $\gamma$-rays (U/Th/K) & 2 & 14.4 \% \\
Lead shield ($^{210}$Pb) & 3.4 & 24.5 \%  \\
Copper shielding (U/Th/K/$^{60}$Co) & 6.6 & 47.6 \% \\
Light guide coating (U/Th/K) & 0.008 & \textless 1 \% \\
PMTs (U/Th/K/$^{60}$Co) & 1.6 & 11.5 \% \\
Vial plastic (U/Th/K) & 0.15 & 1.1 \% \\
Cross-talk (Presumed 2615~keV $\gamma$-ray) & 0.03 & \textless 1 \% \\ 
Neutrons & 0.11 & \textless 1 \% \\ \hline
Total estimated background rate &  13.9 & \\ \hline \hline
\end{tabular}
\end{center}
\end{table}

\section{Evaluation of sample containment}

Due to the elevated concerns for contamination incidents in the ultra-clean shallow underground laboratory, the operations plan is to rely upon double containment of all samples. One method of double containment is to use a 6~mL vial to contain the sample and seal the 6~mL vial within a larger 20~mL vial. Using small sample vials was evaluated in measurements of $^{14}$C with the benefit of producing a lower background as the smaller cocktail volume presents a smaller target for interaction from ubiquitous radiations (\cite{undergrndquantulus2004}). A laboratory measurement comparison of the scintillation emission from the 6~mL vials to the 20~mL vials was performed. As the 6~mL vial is physically contained within a 20~mL vial, a variety of refractive index matching media were included in the interstitial space between the outer wall of the 6~mL vial and the inner wall of the 20~mL vial. The performance was evaluated based upon the collected spectral response of scintillation light in response to the test samples.

\begin{figure}[ht!]
\begin{center}
\includegraphics[width=\columnwidth]{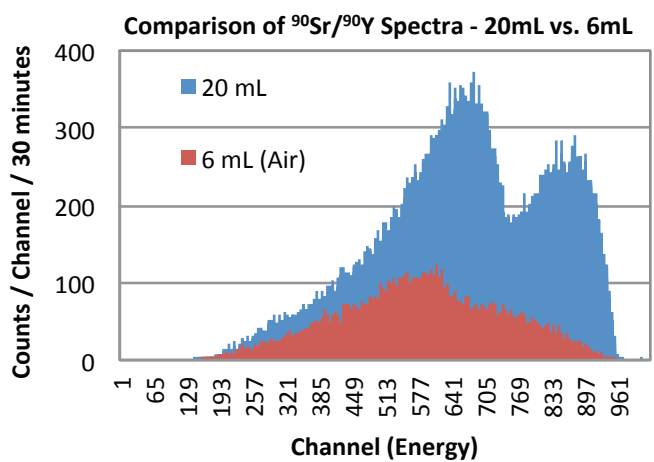}
\caption{\label{fig:Sr90Y90VialSpectra} Spectra for the $^{90}$Sr/$^{90}$Y system taken with a Quantulus system in two vial configurations. The higher (blue) double peak spectrum is from a sample and cocktail within a 20~mL vial. The lower (red) single broad bump spectrum is from a sample and cocktail within a 6~mL vial. The event rates are the same once the ratio of the volumes are taken into account.}
\end{center}
\end{figure}

Using the Quantulus\texttrademark Liquid Scintillation Spectrometer this double containment method was investigated by taking measurements of the spectra of the high energy beta decay of $^{90}$Sr and daughter $^{90}$Y. The vials used were 20~mL Perkin Elmer Superpolyethylene Vials containing a 6~mL Perkin Elmer Pico Prias vial. Within the 6~mL vial was 5.9~mL of the liquid scintillation cocktail Perkin Elmer Ultima Gold AB. Added to this cocktail was 0.07403~mL of a $^{90}$Sr standard, giving the solution an activity of $593 \pm 1$\% decays per minute (dpm). %%593.14$\pm$1.2\%
The vials were sealed and put into the 20~mL vials. Each vial was placed into a different media: air, water, alcohol, and glycol. The four vials were compared to a 20~mL vial that was made with 19.75~mL of Ultima Gold AB cocktail and 0.2527~mL of $^{90}$Sr source with an activity of $2025 \pm 1$\% dpm. %%2024.65$\pm$1.2\%.
The presence of $^{90}$Y in secular equilibrium doubles the total activity rate in the test samples.

\begin{figure}[ht!]
\begin{center}
\includegraphics[width=\columnwidth]{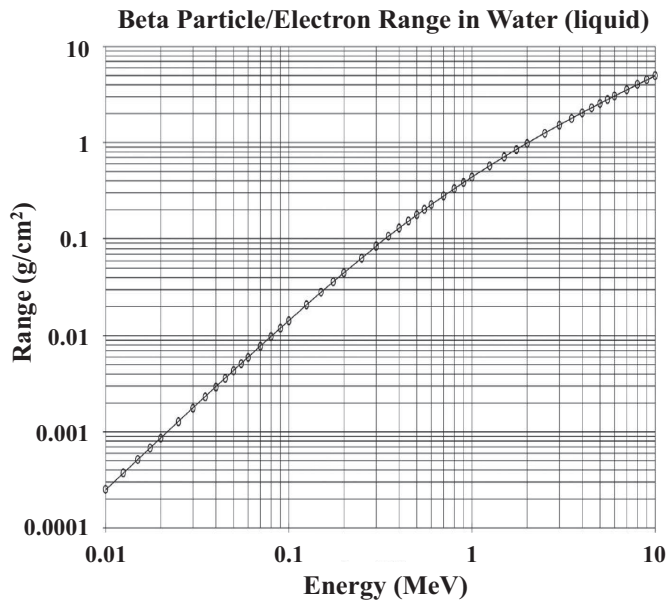}
\caption{\label{fig:BetaInWater}Range of $\beta$-particles in water given in g/cm$^{2}$. The linear distance is determined via dividing by the density of water at 1 g/cm$^{3}$. Figure taken from Appendix B of (\cite{AppB})}
\end{center}
\end{figure}

The outside media (e.g., air, water, alcohol, and glycol) had no effect on the shape of the beta spectrum so only data from the 6~mL vial with air as the surrounding media is used in reporting the following results.  In Figure~\ref{fig:Sr90Y90VialSpectra} the shape of the scintillation spectrum from the 6~mL vial is not comparable to the spectrum from the 20~mL vial exemplar. The 20~mL spectrum distinctively shows two peaks -- the leftmost, lower energy peak corresponding to the $^{90}$Sr and the rightmost, higher energy peak from the $^{90}$Y decay. The scintillation spectrum from the 6~mL vial is degraded and lacks the tell-tale spectral signatures desired for isotopic identification in the $^{90}$Sr/$^{90}$Y system. The integral counts of the two spectra in Figure~\ref{fig:Sr90Y90VialSpectra} are 122,493 counts for the 20~mL sample and 36,197 counts for the 6~mL. These values differ as expected from the differing amount of sample source activity in each vial. In Table~\ref{tab:Sr90Y90Counts} the source activity and volumes are used to show the counting efficiency is equivalent between the two situations.

\begin{table}[ht!]
\begin{center}
\caption{\label{tab:Sr90Y90Counts}
The total integral spectral counts reported from the 6~mL and 20~mL vials -- as seen in Fig.~\ref{fig:Sr90Y90VialSpectra} -- are in agreement once the volume ratio is taken into account. That is, (4137~$\pm$~12)~cpm $\times$ (6~mL / 20~mL) is roughly consistent with (1212~$\pm$~6)~cpm.  Error on the activity is due the knowledge of the activity concentration of the source solution. Error on the measured counts is propagated uncertainty associated with a 30 minute counting duration and $\sqrt{N}$ statistics. No background subtraction was performed (producing efficiency values \textgreater100\%).}
\begin{tabular}{c|c@{~$\pm$~}c|c@{~$\pm$~}c|c@{~$\pm$~}c} \hline \hline
Vial & \multicolumn{2}{c|}{Activity} & \multicolumn{2}{c|}{Counts} & \multicolumn{2}{c}{Efficiency} \\
(mL) & \multicolumn{2}{c|}{(dpm)} & \multicolumn{2}{c|}{(cpm)} & \multicolumn{2}{c}{(\%)} \\ \hline
20 & 4049 & 49 & 4137 & 12 & 102.2 & 1.3 \\
6 & 1186 & 14 & 1212 & 6 & 102.2 & 1.3 \\ \hline \hline
\end{tabular}
\end{center}
\end{table}

Further consideration led to the conclusion the 6~mL vial is volumetrically small relative to the average range of the $\beta$ particles in the scintillation cocktail solution. This range-based effect is apparent upon inspection of Fig.~\ref{fig:BetaInWater}. Note that a 6~mL vial has an inner diameter of approximately 1.5~cm while a 20~mL vial has an inner diameter of approximately 2.5~cm. Both vial sizes have inner heights of approximately 5.0~cm. Table~\ref{tab:BetaRange} shows the approximate distances traveled by $\beta$-particles in water for different energies, produced by different isotopes. As a result of the range of the $\beta$-particle being comparable to the size of the LSC vials, reducing the cocktail volume results in a greater fraction of $\beta$ particles striking the walls of the vial which causes a failure to generate a ``full'' scintillation light emission spectrum relative to the large volume case. This effect is more pronounced for higher energy $\beta$ particles placed in smaller vials. This conclusion is supported by noting both the correct counting rate between the two volumes and roughly the same upper extent of the end of the spectra seen from each vial (near energy channel 961). This effect is not expected to impact tritium ($^{3}$H) measurements as tritium has a much lower, 18~keV beta end-point energy. In conclusion, larger vials will allow for better spectroscopic performance for samples and standardization. This result demonstrates the need to carefully consider how the shallow underground laboratory's procedural requirements for double or triple containment will impact the ultimate performance of the LSC system.

\begin{table}[ht!]
\begin{center}
\caption{\label{tab:BetaRange} Estimated distance of $\beta$-particles in water (having a density similar to LSC cocktails) from consultation of Fig.~\ref{fig:BetaInWater}. Distances are presented for energies relevant to the emissions of several isotopes often measured with liquid scintillation counting.}
\begin{tabular}{c|c@{~$\rightarrow$~}c|c@{~$\rightarrow$~}c} \hline \hline
Isotope & $E_{\beta\mathrm{~End~point}}$ & Distance & $E_{\beta\mathrm{~Average}}$ & Distance \\
            & \multicolumn{1}{c}{(MeV)\phantom{a}}  & \multicolumn{1}{c}{(cm)\phantom{i}} & \multicolumn{1}{|c}{(MeV)\phantom{a}} & \multicolumn{1}{c}{(cm)\phantom{i}} \\  \hline
$^{90}$Y  & 2.280 & 1 & 0.934 & 0.4  \\
$^{90}$Sr & 0.546 & 0.2 & 0.196 & 0.04  \\
$^{14}$C & 0.156 & 0.03 & 0.049 & 0.004  \\
$^{3}$H   & 0.018 & 0.0008 & 0.006 & 0.0002  \\ \hline \hline
\end{tabular}
\end{center}
\end{table}

\section{Future work: Intrinsic and systematic backgrounds}
\label{sec:FutureWork}

The shield design described in this report remains under development while adjustments are made to particular aspects of the final design. The finalized design will inform a high-fidelity GEANT4 simulation of the layout. A U/Th assay of the OFHC copper purchased for the system will better inform the expected background rate as the concentration of trace residual naturally-occurring radioactive decay chain daughters in the copper liner currently dominates the anticipated system background rate. These U/Th assay measurements will also evaluate the actual background contribution level from the reflective coating of the light guides. Measurement of the $^{210}$Pb concentration levels in the lead selected for the shielding will confirm the background contribution. This program of radiochemical assay of the shield assembly materials will provide a foundation for detailed background modeling of the actual intrinsic background levels of the system.

One important systematic background is related to the PMT performance, most notably the dark current rate. This systematic background was not included in the background summary (Tab.~\ref{tab:backgroundsum}) as it is a background \emph{not} related to residual radioactive contaminants in the shield design. The predicted dark count rate for the PMT detectors operated in coincidence is minor compared to the shielding material sources of background -- $\sim$0.69~cpd for a 40~ns coincidence window with a dark count rate of 10~Hz at an equivalent of two or more photoelectrons converted in the PMTs.

There are a number of background phenomenon associated with the scintillation cocktail itself. These include photoluminescence and chemiluminescence (\cite{chapter7}) as well as and U/Th impurities in the scintillation cocktail. Photoluminescence is addressed by not exposing the LSC cocktail to light. The long count duration expected for the ULB LSC system will provide ample time to evaluate any variation in rate due to photoluminescence. Chemiluminescence is additionally addressed through cooling the LSC cocktail. Although not high-lighted in this article, the ULB LSC shield design includes the ability to cool the copper light-guides to 5~degrees Celsius within the nitrogen purged shield. Sample vials will be chilled in a refrigerator prior to introduction to the counting system. Incidentally, the cooling should also provide a small benefit in the reduction of the dark current rates of the PMTs. It is anticipated that the ULB LSC system will find some (but hopefully not all) commercially available liquid scintillation cocktails are contaminated with U/Th/K radioisotopes at the level of $\sim$100 background counts per day. This background rate would not be observable in system with a background rate of $\sim$ count per minute. For ultimate sensitivity and for evaluation of the ULB LSC system, ultra-pure scintillation liquids can be obtained that have levels of U/Th at the $10^{-16}$~g/g level, for example (\cite{BorexinoScint}).

Component testing of the electronics and hardware is underway. Copper light guides will be tested to quantify the light collection performance of the hollow light guides. For comparison and testing purposes the reflective hollow copper light guides' performance will be compared to acrylic light guides in both straight and bent configurations. Testing the hollow guides will show any design issues in creating a seamless surface. For both the copper and lead, the machined pieces must be cleaned before taken into the shallow underground laboratory for assembly.

Construction of the low background liquid scintillator system shield will occur in the summer/fall of 2015 and will entail careful consideration of material cleaning and preparation prior to set-up in the underground laboratory. Data acquisition development to provide $\alpha$/$\beta$ discrimination will occur through the shield construction period. The impacts on background reduction for measurements including both $\alpha$- and $\beta$-emitting nuclei was not considered in this report and is saved for future reports on specific measurements.

\section{Conclusions}
\label{Conclusions}

The design and background simulation results for a low background liquid scintillation counter under development at Pacific Northwest National Laboratory are outlined in this paper. The background contributors analyzed include external gamma rays, radioactivity in the lead and copper shields, and radioactivity in the reflective coating, PMTs, and vials. The copper shielding was the largest background contributor, with the lead shield and reflective coating around the same levels. Characterization of the neutron background will be conducted as part of the future work, but is expected to be small compared to the presented contributors. The background from the natural radioactivity in the PMT glass and ceramics was reduced by bent U-shaped reflective hollow light guides to distance the PMTs from the sample in combination with a lead brick inserted into the line-of-sight from the PMT to the sample. The background estimate from current simulation results is 14 counts per day in the left and right sample vials. By using simulations to inform the design steps, each background contributor investigated is better addressed in the final shield design intended for ultra low-background measurements.

\section{Acknowledgments}
\label{sec:Acknowledgments}

The research described in the paper is part of the Ultra-Sensitive Nuclear Measurements (USNM) Initiative at Pacific Northwest national Laboratory. It was conducted under the Laboratory Directed Research and Development Program at the Pacific Northwest National Laboratory, a multiprogram national laboratory operated by Battelle for the U.S. Department of Energy.





\end{document}